\documentclass[letter]{aa}  

\usepackage{graphicx}
\usepackage[varg]{txfonts}

\begin{document}

   \title{The statistical distribution of magnetic field strength in G-band bright points}


   \author{S. Criscuoli
          \inst{1}
          \and
          H. Uitenbroek\inst{1}
          }

   \institute{ National Solar Observatory, Sacramento Peak, P.O. Box 62, Sunspot, NM 88349, USA
             \email{scriscuo@nso.edu}
             }

   \date{Received ...; accepted ....}

 
  \abstract
   {G-band bright points are small-sized features characterized by high photometric contrast. Theoretical investigations indicate that these features have
   associated magnetic field strengths between 1-2 kG. Results from observations instead lead to
contradictory results, indicating magnetic fields of only kG strength in some and including hG strengths in others.}
   { In order to understand the differences between measurements reported in the literature, and to reconcile them with results from theory,
   we analyze the distribution of magnetic field strength 
   of G-band bright features identified on synthetic images of the solar photosphere,
 and its sensitivity to observational and methodological effects.   }
   {
We investigate the dependence of magnetic field strength distributions
of G-band bright points identified in 3D magnetohydrodynamic simulations on feature selection method, data sampling, alignment and  spatial resolution. 
}
   {The distribution of magnetic field strength of G-band bright features shows two peaks, one at about 1.5 kG and one below 1 hG. The former corresponds to magnetic features, 
   the second mostly to bright granules.
   Peaks at several hG are obtained only on spatially degraded or misalligned data.     }
{Simulations show that magnetic G-band bright points have typically associated field strengths of few kG. Field strengths in the hG range can result from observational 
effects, thus explaining the discrepancies presented in the literature.
  Our results also indicate that outcomes from spectro-polarimetric inversions with
   imposed unit filling-factor should be employed with great caution.    }

   \maketitle
%

\section{Introduction}
\label{intro}
G-band Bright Points (BPs, hereafter)  are roundish features, a few hundred kilometers in diameter, whose contrast with respect to quiet
regions is high (usually 30\% or more) when observed in Fraunhofer's G-band (the spectral range of about 1 nm around 430.5 nm).
Co-temporal and co-spatial observations with magnetograms show that some BPs have associated magnetic 
flux concentrations, while other correspond to bright granules (Keller \cite{keller1992}; Berger \& Title \cite{berger2001}; 
de Wijn et al. \cite{deWijn2009} for a review); the two populations also present different spectral characteristics in 
the G-band (Langhans, Schimdt \& Tritschler \cite{langhans2002}).

Theoretical studies have shown that the brightening of magnetic features in the G-band is due to the weakening of CH molecule lines (which are conspicuous in this Fraunhofer band), 
which results from the shallower temperature and reduced pressure and density within magnetic structures with respect to the surrounding quiet regions 
(e.g. Sch\"{u}ssler et al. \cite{schussler2003}, Uitenbroek \& Tritschler \cite{uitenbroek2006}).
Results obtained by Sch\"{u}ssler et al. (\cite{schussler2003}) and Shelyag et al. (\cite{shelyag2004}) from the analysis of
magnetohydrodynamic (MHD) simulations indicate that such conditions are satisfied only in kG structures.

Some observations confirm the kG nature of G-band BPs. For instance, Ishikawa et al. (\cite{ishikawa2007}), who analyzed data from the Swedish 
Solar Telescope (SST), retrieved an average magnetic flux of
$\approx$ 1.5kG. Viticchi\'{e} et al. (\cite{viticchie2010})  also found a field strength of 1.5 kG by inverting spectro-polarimetric data acquired 
at the Dunn Solar Telescope (DST). However, Beck et al. (\cite{beck2007}), from spectro-polarimetric inversions of 
data acquired at the Vacuum Tower Telescope (VTT), and of
simultaneous G-band observations from the Dutch Open Telescope (DOT), deduced a rather flat field strength
distribution ranging between $\approx$  0.5 kG and 1.5 kG. More recently, Utz et al. (\cite{utz2013}), by the analysis  of BFI/HINODE G-band data and spectro-polarimetric 
inversions of SP/HINODE data, found that the magnetic field distribution of BPs can be described by the superposition 
of four log-normal functions, two of which have peaks in the 
kG range, and two in the hG range. These authors concluded that features in the kG range correspond to "collapsed fields" (subdivided into 
"weak collapsed field", with a peak at $\approx$ 1.1 kG, and "strong collapsed field", with a peak at $\approx$1.3 kG); features whose field distribution 
peaks at $\approx$ 7 hG correspond to "pre- and post- collapsed" magnetic field; and features whose field distribution peaks at $\approx$ 3 hG correspond to
"background" field related to solar dynamo processes. It is worth to note that while Utz et al. (\cite{utz2013}) employed results from spectro-polarimetric 
inversion with an imposed unit filling factor, Beck et al. (\cite{beck2007}) and Viticchi\'{e} et al. (\cite{viticchie2010}) assumed two component atmospheres, 
one magnetic and one quiet, so that the filling factor was a free parameter. 
  
Due to their small-size, which is still at the limit of the spatial resolution of modern instrumentation, the estimation of properties of BPs is 
prone to observational effects as has been showed by numerical models (e.g. Criscuoli \& Rast \cite{criscuolirast2009}) and 
observations (e.g. Viticchi\'{e} et al. \cite{viticchie2010}). We therefore want to qualitatively investigate whether the discrepancy of results presented in the 
literature can be attributed to differences in the  quality of the data and in the tools employed for their analysis.
With this aim we investigated the effects of image degradation, image-thresholding, pixelization and  misalignment between G-band and magnetograms, of
magnetic field distributions of bright features identified on G-band images derived from 3D MHD simulations.
The paper is organized as follows: in Sec. 2 we describe the simulations and their analysis; in Sec. 3 we present our results and in Sec. 4 we draw our conclusions.

   \begin{figure*}
   \centering
   \includegraphics[width=4.5cm, trim=0.5cm 0cm 5cm 0.5cm]{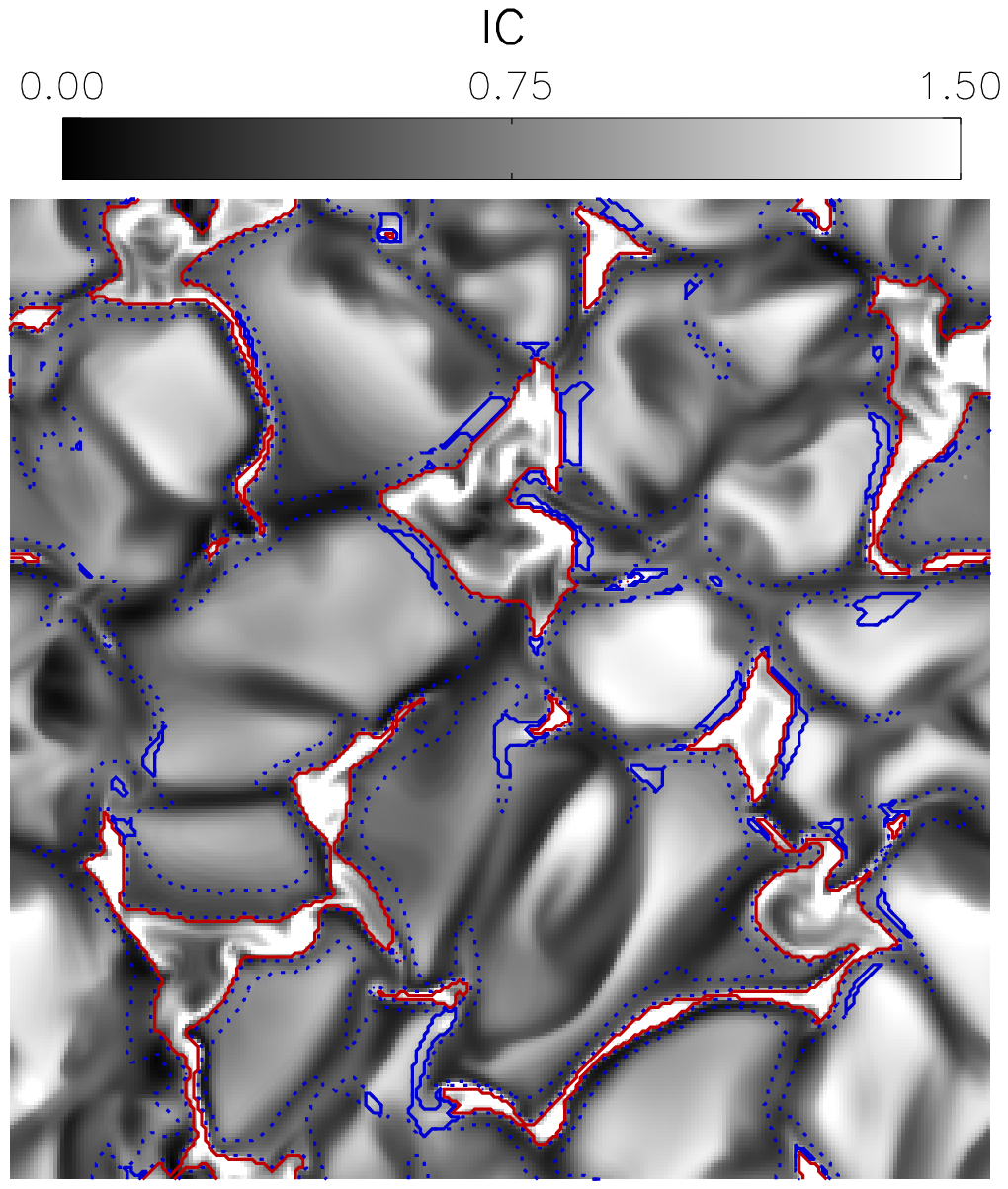}
      \includegraphics[width=4.5cm, trim=0.5cm 0cm 5cm 0.5cm]{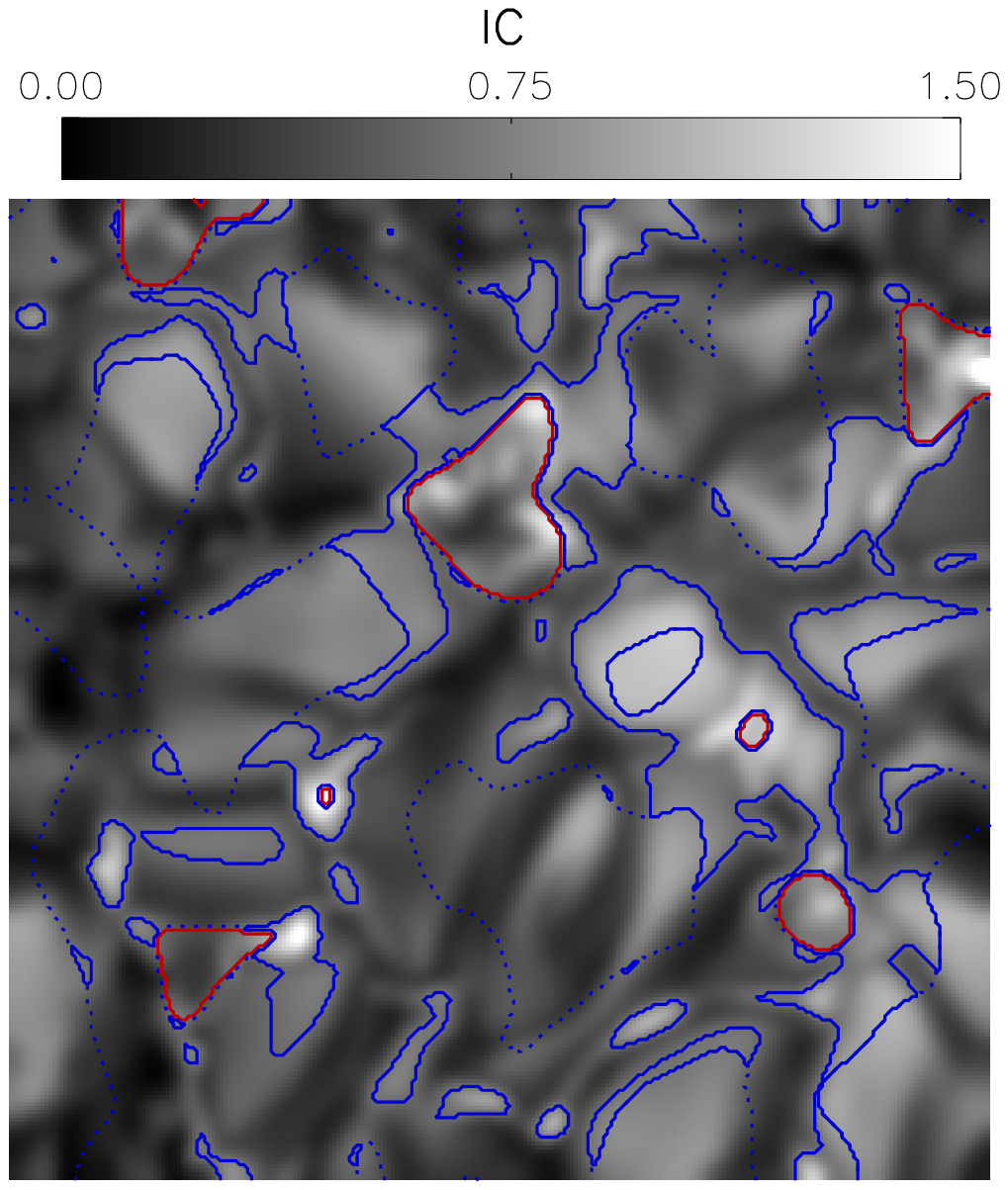}
   \includegraphics[width=4.5cm, trim=0.5cm 0cm 5cm 0.5cm]{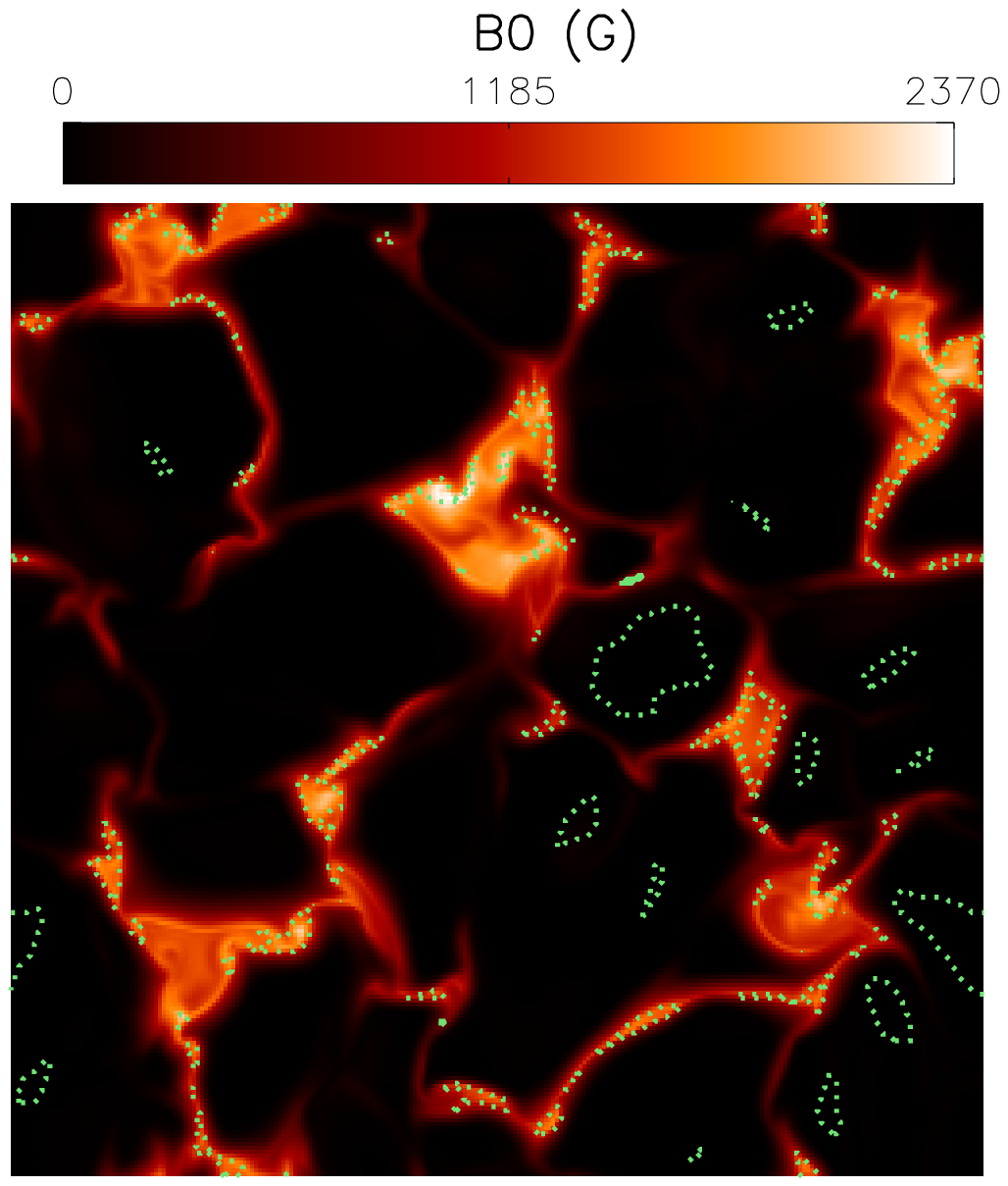}
      \includegraphics[width=4.5cm, trim=0.5cm 0cm 5cm 0.5cm]{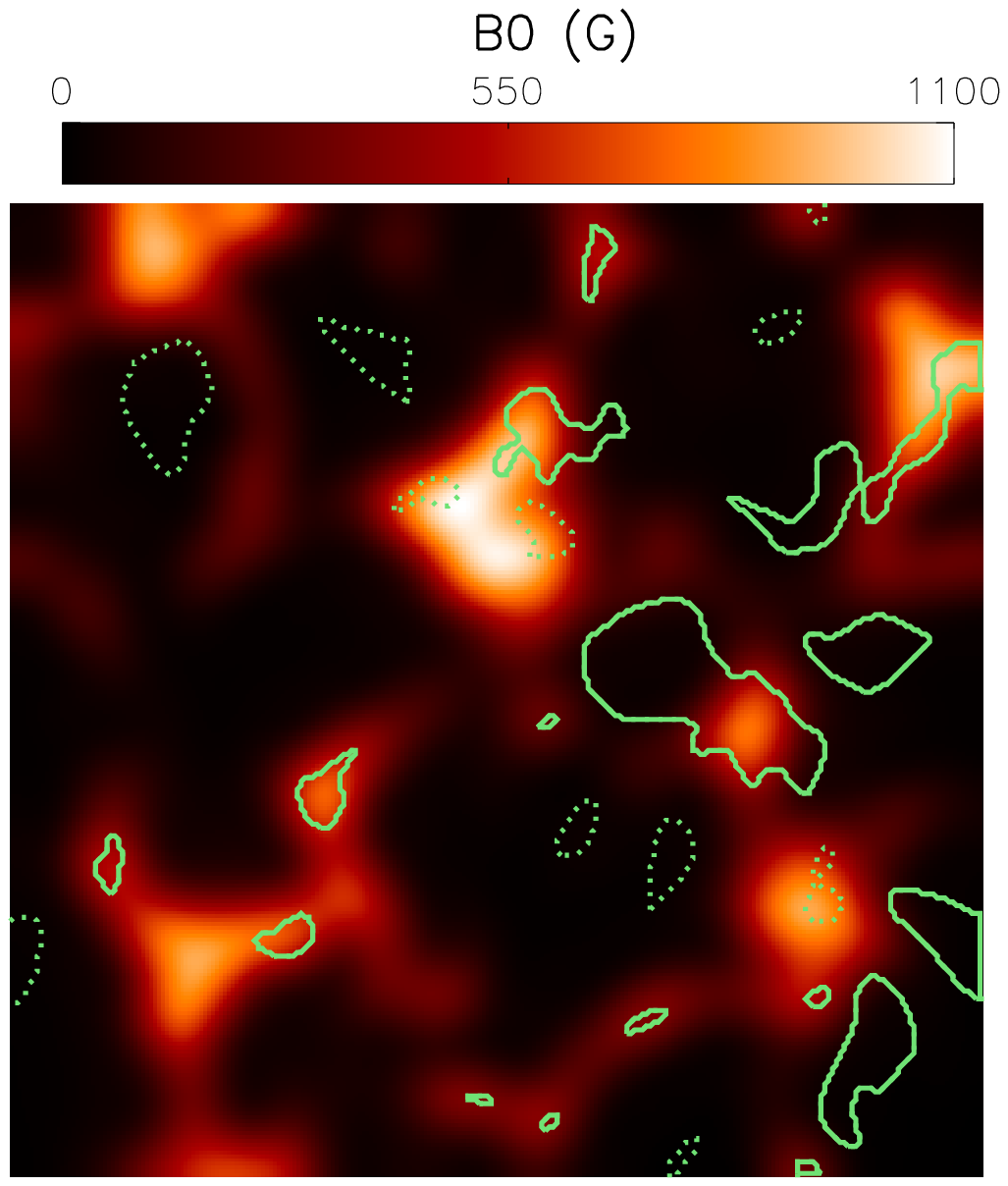}
  
      \caption{Left: G-band contrast IC of one of the MHD snapshots, saturated at 1.5. Blue dotted lines: borders of regions where 1hG$<$B0$<$1kG. Blue solid lines:
      regions where 1hG$<$B0$<$1kG and IC $\geq$1. Red solid lines: regions where B0$\geq$1kG. Middle-Left: same as left panel, 
      but image is degraded with PSF\_50cm\_430.
      Middle-Right: B0 corresponding to left panel. Right: B0  convolved with 'PSF\_50cm\_06'. Green dotted lines denote
      contours of features identified on images in left and middle-left panels, respectively,  assuming $\alpha$=1.5; 
      green continuous lines denote borders of a sub-set of identified features whose average field value is between 1hG and 1kG. 
  }  
              \label{FigImages}
   \end{figure*}
 
  
\section{MHD simulations and data analysis}
We employed nine snapshots from a 3D MHD simulation covering an area of 6$\times$6 Mm$^2$ of the solar photosphere obtained with the Copenhagen-Stagger code 
(Nordlund \& Galsgaard \cite{nordlund1995}), characterized by having 2hG average magnetic
flux and spatial sampling of $\approx$ 0.03``/pixel in the horizontal direction (see
Fabbian et al. \cite{fabbian2010} and Fabbian et al. \cite{fabbian2012} for a detailed description). Since properties of magnetic features are known to 
vary on the magnetic flux of their environment (Criscuoli \cite{criscuoli2013} and references therein), we also considerd snapshots having 0.5 and 1 hG average magnetic flux.
Nevertheless, since  results obtained from these latter simulations are similar to those obtained from 2 hG simulations, but present lower statistics,   
in the following we show only results obtained from the 2 hG simulations. 
The snapshots were randomly selected with the constraint of being between 6 and 11 minutes apart; this sampling ensured that 
the snapshots were independent from  each other, and also reduced effects introduced by p-modes.
Each snapshot was spatially resampled in the vertical dimension as described in (Criscuoli \cite{criscuoli2013}), and the G-band spectrum was synthesized in the 
vertical direction with 
the RH code (Uitenbroek \cite{uitenbroek2002}, Uitenbroek \cite{uitenbroek2003}) as described in Uitenbroek,
Tritschler \& Rimmele \cite{uitenbtrit2007}.  
Intensity images were then obtained by multiplying the spectra by a Lorentian-shape filter profile of Full Width at Half Maximum equal to 1 nm and centered at 430.5 nm.
Note that we found that variations of up to 40\% in the width of
the filter do not produce significant variations of the results presented below. 
 We also synthesized the intensity in the continuum at 630 nm.
 
Bright features were identified on each snapshot by considering all those pixels whose G-band intensity contrast
IC (defined as the ratio between the G-band intensity of the pixels and the median intensity in each snapshot) satisfied the relation
IC $\ge$ M + $\alpha * \sigma $, where $M$ and $\sigma$ are the median and standard deviation of the contrast, respectively,  
within the snapshot and $\alpha$ is a free parameter
that we let vary between 0.5 and 1.5. Each feature was then labeled with the IDL 'label\_region' routine. The number of features identified in
this way in original data varied between 730 to 800, depending on the $\alpha$ values,
and varied between 300 and 350 in spatially degraded data (see text below).  We then produced maps of the magnetic
field strength at optical depth $\tau_{500}$ = 0.1 (B0, hereafter) and for each identified feature we considered 
their average field strength and their field strength at 
the pixel corresponding to the G-band intensity barycenter over the B0 maps. Since we obtained similar distributions and trends with the two methods, 
in the following we only present results obtained for average magnetic field strengths. Note that Orozco Su\'{a}rez et al. \cite{orozco2010} showed that Milne Eddington
inversions statistically provide reliable estimates of magnetic field properties at $\tau_{500}$ = 0.1, therefore the magnetic field strength values computed over B0 maps are
good representations of results obtained from inversions.

In order to investigate the dependence of the distributions on spatial resolution, we convolved the intensity images, as well as  B0 maps, 
with Point Spread Functions (PSFs, hereafter) of different shapes. Following Wedemeyer-B\"{o}hm (\cite{wedemeyer2008}), 
we modeled the PSFs as the convolution between an Airy function, which is the PSF of telescope aperture,
and a Voigt function, which takes scattered-light into account. 
The free parameters of the PSF model were varied to represent five different cases. Two were obtained assuming a telescope aperture of 50 cm and wavelengths of
630 nm and 430 nm, respectively, which determine the parameters of the Airy functions. We set the parameters for the Voigt function to reproduce an
average rms contrast over the snapshots similar to those derived from HINODE BFI and SP measurements, i.e., 0.11\% for the G-band (Mathew et al. \cite{mathew2009})
and 7\%  for the 630 nm images (Danilovic et al. \cite{danilovic2008}). In the following we will refer to these functions as 'PSF\_50cm\_430' and 'PSF\_50cm\_630'.
 They approximately represent the PSFs of HINODE BFI and SP. Following results obtained by 
Beck et al. (\cite{beck2013}), to represent the PSF of SP we also constructed a third function 
whose Full Width at Half Maximum is 0.6`` and whose wings amplitude  is similar to that
obtained by Beck et al. (\cite{beck2013}) ('PSF\_50cm\_06', hereafter); the average rms contrast in our snapshots using this PSF is about 6\% at 630 nm. 
Two other PSFs, representing a telescope aperture of 70 cm at 630 nm and 430 nm were also calculated ('PSF\_70cm\_630' and 'PSF\_70cm\_430', hereafter); these approximately 
represent the PSFs of the VTT and the DST. In order to 
investigate the effects due only to the spatial resolution,
we kept the free parameters of the Voigt functions equal to those derived for the 50cm PSFs. We also tested varying the wing amplitudes of the PSFs; obtained results 
are briefly discussed below.

Finally, we analyzed the dependence of the results on spatial sampling, rebinning the data to one and two thirds of their original sizes,  
and on the misalignment between the G-band images and B0 
maps by shifting these data with respect to one another.
 
 \section{Results}
Comparing G-band contrast and B0 maps we find, in general, that hG fields
mostly occur in intergranular lanes, while kG fields preferentially
appear at vertexes between granules with some presence in lanes.
However, we also find that only the kG features are bright in the G band.
The contours of the two classes of features
 are marked in dotted blue and solid red in the left image of Fig.\ref{FigImages}.  
     \begin{figure}
   \centering
        \includegraphics[width=6.5cm, trim=2.cm 1cm 1cm 1cm]{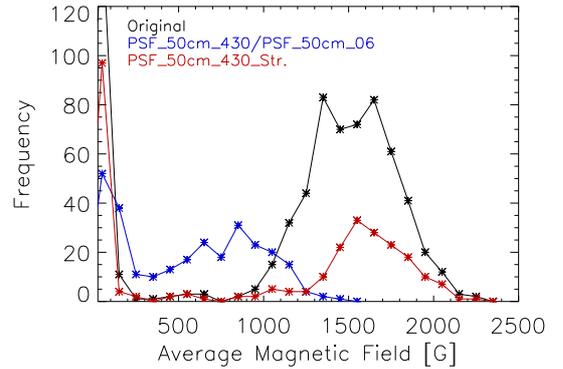}
            \caption{ Statistical distribution of the average magnetic field of features identified assuming $\alpha$=1.5 in G-band and B0 images (black symbols),
            both degraded (blue symbols), and in the case where only G-band images are degraded (red symbols). Bin size = 100G. The PSFs employed to degrade
            the data are indicated in the labels.}  
              \label{plot1}
   \end{figure}
 The solid blue lines also marked in the same panel represent pixels for which 1 hG$<$B0$<$1 kG and IC$>$1; these are 
  few and are mostly located at the borders of larger magnetic regions. Inspection of data shows that they
 result mostly from the expansion of the field with height, while a small fraction has associated horizontal field only. On degraded images the area
 occupied by positive-contrast hG pixels is larger (middle-left panel). 
 
 Middle-right and right panels show examples of contours of features identified assuming $\alpha$=1.5 on original and 
 spatially degraded data, super imposed on  B0 maps,  and B0 maps degraded with PSF\_50cm\_630, respectively.   Clearly, 
  two types of features 
 are identified on both types of data:  bright granules (for which the field strength is $<$ 1 hG) and kG features. 
 Only a small amount of features identified on non-degraded data have hG field strength (for instance, on the middle right image only one of such features is identified); 
 these usually encompass bright granules adjacent to small-size magnetic patches. 
 On spatially-degraded data, the number of identified features having hG field strength is larger. These  in part correspond to bright granules close to features that, on the original images,
 had kG field, and in part
 correspond to small-size features that had associated kG field in the original data, and whose average field decreased as an effect of the spatial degradation.

 Magnetic field distributions of the identified features are illustrated in Fig. \ref{plot1}.
 The black symbols represent the distribution of features identified on the original G-band intensity images. We notice a peak at $\approx$ 1.5 kG, 
 with values up to 2 kG; the
  small tail at
 hG field strengths confirms  that features identified on non-degraded images whose average field strength is of some hG are statistically irrelevant.
 The blue symbols denote the distribution of features identified on G-band images degraded with PSF\_50cm\_430 and the 
 corresponding B0 degraded with PSF\_50cm\_06; in this case the distribution covers a wider range of average field strength values, 
 and the peak occurs at $\approx$ 
 7hG.
 The red symbols show the distribution of features identified on the
  G-band snapshots degraded with 'PSF\_50cm\_430', but B0 non-degraded. This latter case would simulate results 
 obtained from a 'perfect' spectro-polarimetric inversion performed with filling-factor as free parameter,  capable of returning the ''real`` 
 magnetic field strength value (these cases are 
 denoted with the suffix '\_Str.' hereafter); indeed, the distribution in this case is more similar to the one obtained from the original snapshots, 
 as the peak occurs at $\approx$ 1.5kG. In all distributions, the highest peak occurs at B0$<$ 1 hG; these correspond to bright granules or to granules adjacent 
 to magnetic features identified as a single feature, as discussed above. 
 
\begin{figure*}
   \centering
       \includegraphics[width=6.cm, trim=2.cm 1cm 0.8cm 1cm]{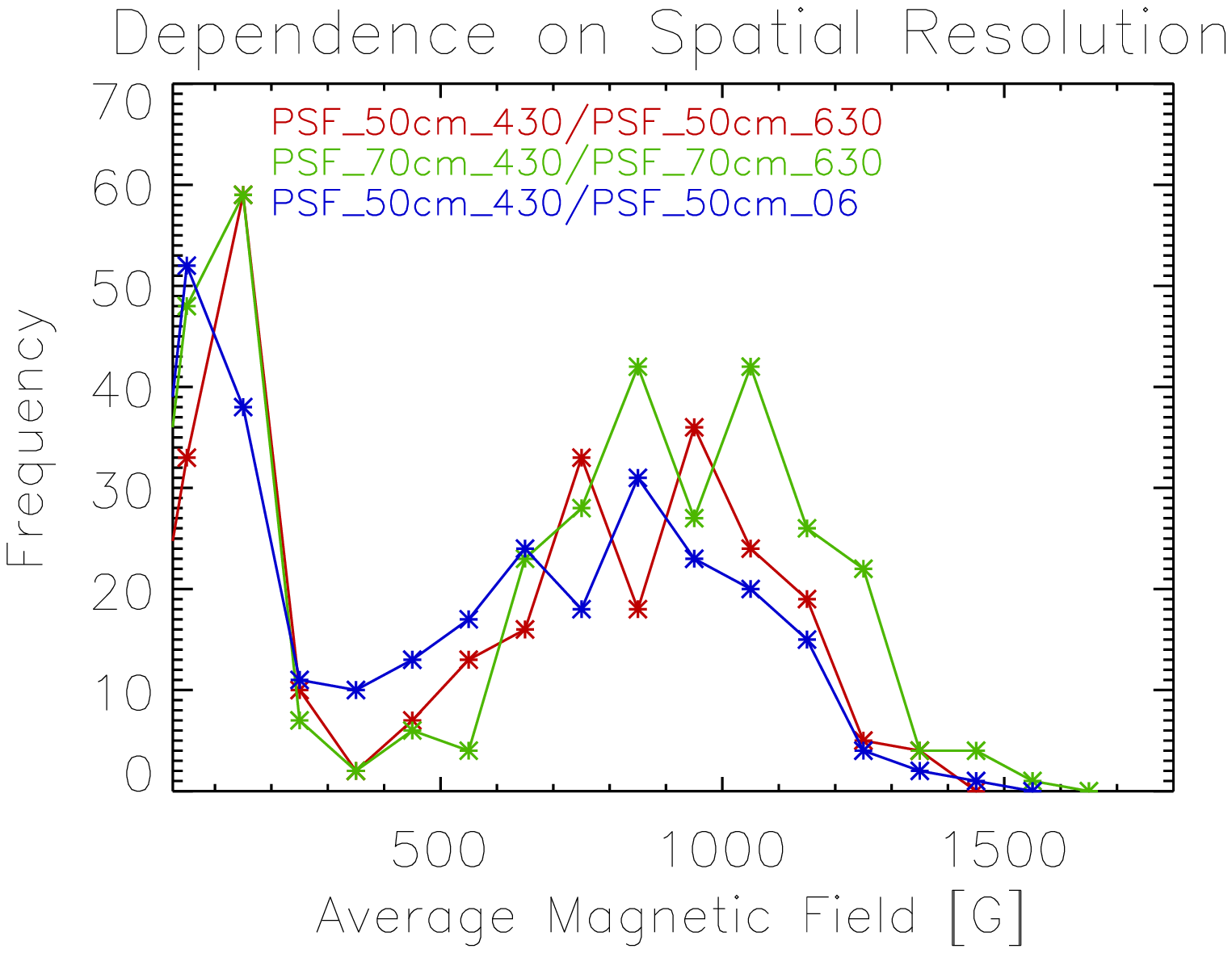}
         \includegraphics[width=6.cm, trim=2.cm 1cm 0.8cm 1cm]{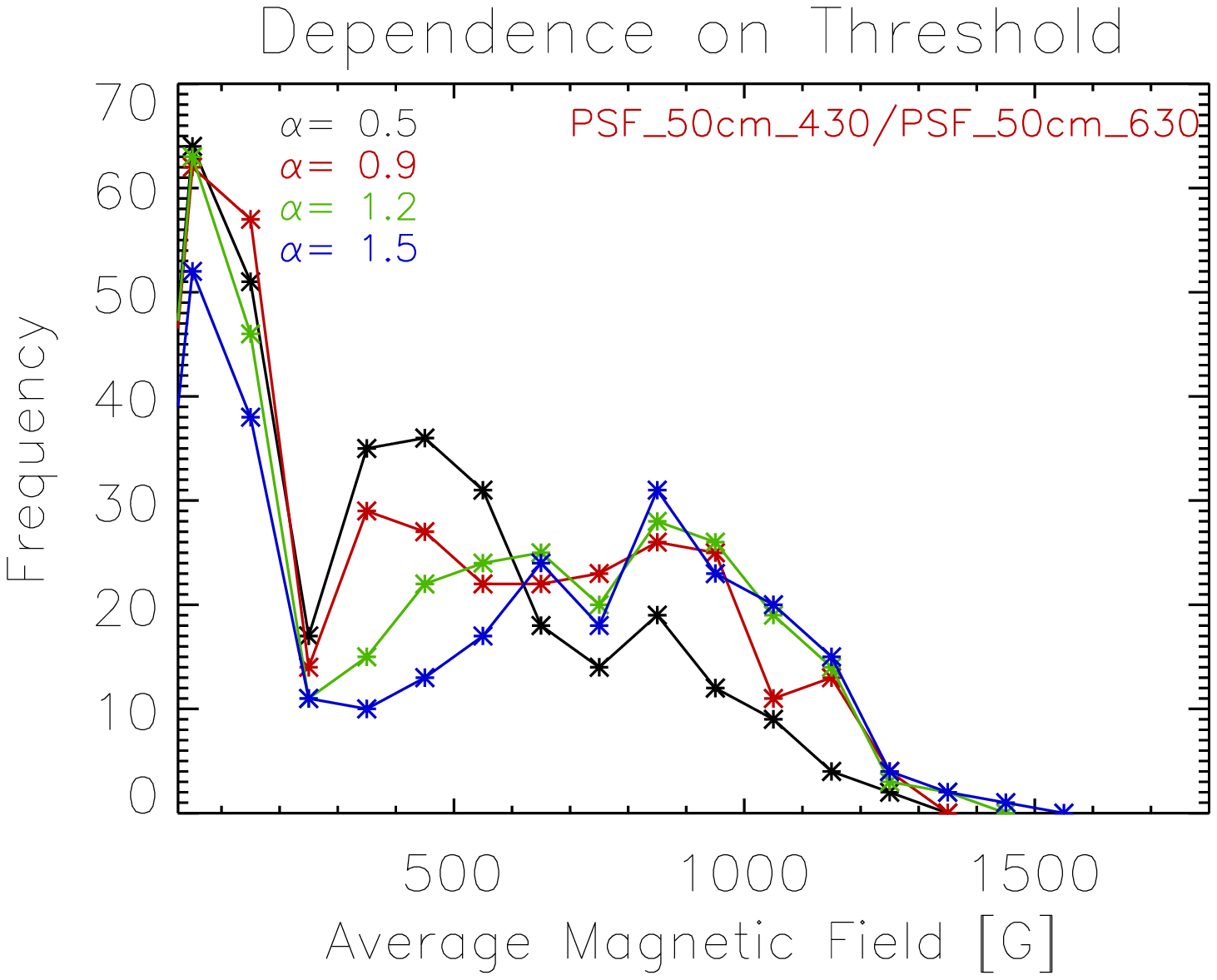}
           \includegraphics[width=6.cm, trim=2.cm 1cm 0.8cm 1cm]{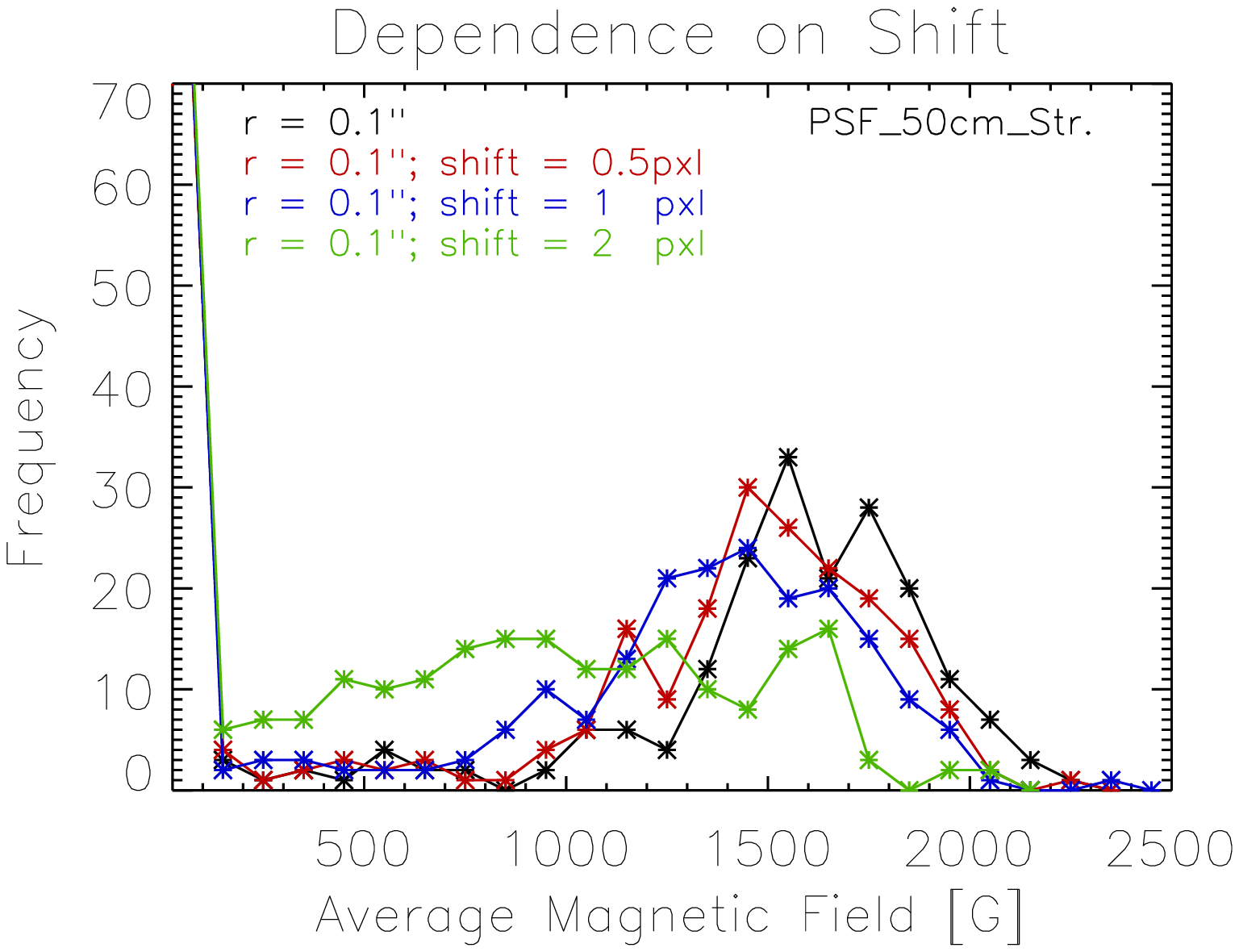}
            \caption{ Dependence of distributions on observational and methodological effects.
      Left: distribution of the average field of features identified assuming $\alpha$=1.5 on G-band images and B0 degraded with different PSFs (see legend). 
      Middle: distribution of the average field of features identified assuming  different $\alpha$ values (see legend), 
      on both G-band and B0 images degraded with the indicated PSFs. Right: distribution of the average magnetic field of features identified assuming 
      $\alpha$=1.5 on G-band images of 0.1'' spatial sampling degraded assuming the indicated PSF, and B0 shifted in the x-direction (see legend).}
            \label{plots}
   \end{figure*}

 We then compared distributions obtained from data convolved with the different PSFs described in Sec. 2.  Left panel in Fig. \ref{plots} shows
 results obtained convolving both G-band images and B0. We notice that the peaks of the distributions shift toward lower values, being $\approx$ 9 hG in the case 
 of a telescope of 70 cm diameter and $\approx$ 7 hG in the case of a 50 cm. Similarly, the widths of the distributions 
 increase with the decrease of the spatial resolution.
 Similar results were obtained when only the G-band images were degraded (not shown).
 
The middle panel in Fig.\ref{plots} shows the effects on magnetic field distributions of image thresholding in the case of 
B0 and G-band images degraded with PSF\_50cm\_06 and  PSF\_50cm\_430, respectively. We notice that peaks shift toward lower values and distributions broaden
with the decrease of the threshold value. Results obtained degrading only G-band images
 show qualitatively the same trends (not shown), with the distributions resulting from $\alpha$=0.5 being rather flat.   
 Note that we found that increasing the amount of scattered-light has qualitatively the same effect as decreasing the threshold value. 
 
 We also investigated the effects of pixel sampling on the shapes of the magnetic field distributions. With this aim, we compared distributions obtained at 
 the original pixel scale of 0.03'' with images resampled at
 approximately 0.05'' and 0.1''. We found that, for both original and degraded images, rescaling does not affect the shape of the distributions, as is shown for instance 
 by comparison  of the distribution obtained from 0.03`` pixel scale data (represented with the blue symbols in Fig. \ref{plot1}), and the distribution obtained
 from 0.1'' pixel/scale data (represented with the black symbols in Fig. \ref{plots}).
 
 Finally, we investigated the effects of misalignment between the G-band images and B0. We found that with the increase of the amount of misalignment 
 the peaks of the  distributions shift toward lower values and their widths increase. Nevertheless, shapes of distributions are significantly 
 altered only for misalignments comparable to or larger than the spatial resolution of the data (i.e. the amount of spatial degradation); therefore, 
original data distributions are more affected by misalignment than distributions obtained
 from degraded data. The plot in right panel of Fig. \ref{plots} represents an intermediate case, as it shows results obtained when only G-band images are degraded with
 PSF\_50cm\_430. Note that the plot shows results obtained on data whose pixel scale is 0.1''. Here, it is also worth to note that even for the intermidiate case in which 
 only G-band images are degraded,  a sub-pixel shift leads to
 a non-negligible modification of the shapes of distributions.

\section{Conclusions}
We implemented an automatic method to select bright features on synthesized G-band intensity images obtained from 3D MHD simulations.
We found that most of the bright features identified on non-degraded 
images correspond to either kG features or to granules. As a result, 
the distribution of the average magnetic field of the identified bright features presents two populations. One spans the range  1 - 2 kG and has 
a peak at about 1.5 kG, the other spans the range 0 - 2 hG. Since we obtained similar results from snapshots characterized by different amounts
of magnetic flux, we conclude that G-band BPs 
harbor kG magnetic field regardless of the region (quiet or active) they are embedded in. We then investigated the sensitivity of the shape of the distribution on data quality and the employed feature selection method.  We found that the distributions widen 
and shift toward lower magnetic field values when images are spatially degraded, when the threshold value employed for identification is lowered and when the 
the misalignment between G-band images and magnetic field data is increased. On the other hand, distributions are rather insensitive to the spatial sampling, 
as far as 
this is good enough to resolve the BPs.

Our results therefore suggest that the different shapes of distributions presented in the literature mostly result from the different identification 
methods employed to select features and the quality of the data analyzed. In particular, we speculate that the broad distribution obtained by Beck et al. 
(\cite{beck2007}) could be the result of residual temporal and spatial misalignment between the data. Results by Utz et al. (\cite{utz2013}), obtained employing
spectropolarimetric
inversions with unit filling factor, are most likely
affected by image degradation. In particular, it is likely that the kG component they found results from larger magnetic structures, which are 
less affected by spatial degradation, on which the inversion could retrieve reliable estimates of the field intensity. The population peaking at 7 hG that they found,
instead, results 
from  smaller-size elements, for which the inversion returned the magnetic flux, not the field strength. Finally the population that peaks at about 3 hG most likely
encompasses bright granules adjacent to magnetic concentrations, whose flux therefore turns to be of a few hG, and patches of horizontal field, 
whose number is known to be underestimated in MHD simulations without local dynamo (e.g. Danilovic et al. \cite{danilovic2010}) like the ones that we employed.
It is also worth to note that the identification method 
employed by Utz et al. (\cite{utz2013}) selects structures that are bright with respect to the local background. As a result, features
such as umbral dots or penumbral and light-bridge features are also selected (see their Fig.4). These features are embedded in the canopy of the pore or 
sunspot they belong to, so that an inversion performed with unit filling factor is more likely to return kG magnetic field strength.
This explains the remarkable difference that these authors
found between distributions obtained in quiet areas and around a pore. Finally, in our simulations we do not find a clear indication of two populations
of bright features in the kG range. In fact,
the distribution obtained on original data reported in Fig.\ref{plot1} might suggest two peaks at about 1.4 kG and at 1.7 kG, but inspection of simulations
 did not
reveal any relevant physical difference between the two populations.

Our results therefore suggest that great care should be taken when employing results from inversions performed with unit filling factor (see also 
Orozco Su\'{a}rez et al. \cite{orozco2007}).

\acknowledgements
The snapshots of magnetoconvection simulations were calculated using the 
computing resources of the MareNostrum (BSC/CNS, Spain)
and DEISA/HLRS (Germany) supercomputer installations.

\end{document}